\documentclass[useAMS,usenatbib]{mn2e}

\usepackage{graphicx,psfrag}

\title[Supermassive Black Holes in Cosmological Simulations]{Modelling the Growth of Supermassive Black Holes in Cosmological Simulations}
\author[S.I. Muldrew et al.]{Stuart I. Muldrew$^{1}$\thanks{E-mail:
stuart.muldrew@nottingham.ac.uk}, Frazer R. Pearce$^{1}$ and Chris Power$^{2}$\\
$^{1}$School of Physics and Astronomy, University of Nottingham, Nottingham, NG7 2RD, UK\\
$^{2}$ICRAR, University of Western Australia, 35 Stirling Highway, Crawley, Western Australia 6009, Australia}
\begin{document}

\date{Accepted ?? ?? ??. Received ?? ?? ??; in original form 2013 June 18}

\pagerange{\pageref{firstpage}--\pageref{lastpage}} \pubyear{2013}

\maketitle

\label{firstpage}

\begin{abstract}
There is strong evidence that supermassive black holes reside in all galaxies that contain a stellar spheroid and their mass is tightly correlated with properties such as stellar bulge mass and velocity dispersion.  There are also strong theoretical arguments that feedback from supermassive black holes plays an important role in shaping the high mass end of the galaxy mass function, hence to accurately model galaxies we also need to model the black holes.  We present a comparison of two black hole growth models implemented within a large-scale, cosmological SPH simulation including star formation and feedback.  One model is a modified Bondi-Hoyle prescription that grows black holes based on the smooth density of local gas, while the other is the recently proposed Accretion Disc Particle (ADP) method.  This model swallows baryonic particles that pass within an accretion radius of the black hole and adds them to a subgrid accretion disc.  Black holes are then grown by material from this disc.  We find that both models can reproduce local scaling relations, although the ADP model is offset from the observed relations at high black hole masses. The total black hole mass density agrees between models to within a factor of three, but both struggle to reproduce the black hole mass function.  The simulated mass functions are too steep and underestimate the number of intermediate and high mass black holes.  In addition, the ADP model swallows excessive amounts of material at the resolution of large-scale, cosmological simulations producing unrealistically large accretion discs.  Future work needs to be performed to improve the black hole mass function within simulations. This should be done through the mass growth and feedback as they are strongly coupled and should not be treated as separate entities.
\end{abstract}


\section{Introduction}
\label{intro}

Supermassive Black Holes (SMBHs) are hosted at the centre of all galaxies with a stellar spheroid \citep{Kormendy95,Ferrarese00} and play an important role in galaxy evolution.  Without the feedback they power through Active Galactic Nuclei (AGN) it is difficult to reconcile the observed high mass end of the galaxy stellar mass function with that predicted by galaxy formation models \citep{Bower06,Croton06}.  In addition there is mounting evidence for the coevolution of galaxies and SMBHs through the SMBH Mass--Spheroid Velocity Dispersion relation \citep[$M_{\rm BH}-\sigma$:][]{Gebhardt00,Tremaine02} and the SMBH Mass--Bulge Stellar Mass relation \citep[$M_{\rm BH}-M_{\rm Bulge}$:][]{Magorrian98,McLure02}.  Molecular outflows have also been observed with mass-loadings far higher than that expected from star formation alone which are likely to be powered by AGN \citep{Feruglio10,Alatalo11}, giving further evidence of their influence in galaxy formation.

The exact mechanism for the formation of SMBHs remains uncertain, but there are three main theories that predict different seed masses.  The first is that massive Population III stars collapse giving black hole seeds of $10^2-10^3\, {\rm M}_{\odot}$ \citep{Madau01}; alternatively the collapse of atomically cooling $\sim10^4\, \rm K$ primordial gas in dark matter haloes may lead to seed masses of $10^4-10^6\, {\rm M}_{\odot}$ \citep{Bromm03}.  The third mechanism is that they may form from the collapse of $\sim10^3\, {\rm M}_{\odot}$ stars created in runaway collisions in dense stellar clusters \citep{Devecchi09}.  \citet{Johnson12} suggest that the lower limit on SMBH seeds is $\sim10^5\, {\rm M}_{\odot}$ which requires significant rapid growth to produce SMBH of $2\times10^9\, {\rm M}_{\odot}$ at $z\sim7$ \citep{Mortlock11} and $1.7\times10^{10}\, {\rm M}_{\odot}$ by the present day \citep{vdB12}.

In the context of cosmological simulations, seed black holes are initially many times smaller than the typical resolution of hydrodynamic particles \citep{Schaye10,DiMatteo12} and the exact details of their physics is too poorly understood to simulate directly.  This results in the formation, growth and feedback of SMBHs being added in a subgrid manner.  Sink particles are used to represent the SMBH with a subgrid accretion scheme implemented \citep*{Springel05a}.  The most common accretion model used in the literature is the Bondi-Hoyle \citep{Bondi44,Bondi52} method \citep[e.g.][]{Springel05a,DiMatteo05,Sijacki07,DiMatteo08,DiMatteo12,Vogelsberger13}.  Bondi-Hoyle models the accretion of a spherically symmetric uniform flow of zero angular momentum material captured gravitationally by a point source.  This results in an accretion rate onto the SMBH, $\dot{M}_{\rm BH}$, that is proportional to the mass of the SMBH squared, the local density of the gas, $\rho$, and inversely proportional to the sound speed, $c_{\rm s}$, cubed, i.e. $\dot{M}_{\rm BH} \propto \dot{M}_{\rm Bondi} \propto M_{\rm BH}^2\rho/c_{\rm s}^3$.

Although commonly used in simulations, the Bondi-Hoyle method has a number of limitations, as discussed in \citet{Hobbs12}.  The principle assumption is that gas is at rest at infinity, but SMBHs are embedded within stellar bulges and dark matter haloes that are many times larger.  If the gas within the halo is as hot as the virial temperature, then it will be in hydrostatic equilibrium and the Bondi-Hoyle method will apply.  However, during periods of rapid growth of the SMBH, the halo is gas rich and dense gas is likely to cool faster.  This will lead to the gas collapsing to the centre triggering star formation and feeding the SMBH.  In this case, there is a net radial inflow towards the SMBH and so gas cannot be assumed to be at rest at infinity and violates the Bondi-Hoyle assumption.

Another assumption of Bondi-Hoyle is that the gas accretes onto the SMBH with zero angular momentum, which is known to be not true.  As gas collapses onto the SMBH it will settle into a circular orbit forming an accretion disc, whose radius is set by the angular momentum of the gas relative to the SMBH.  This angular momentum forms a natural barrier to accretion and only low angular momentum gas will be accreted onto the SMBH \citep{King10,Hobbs11}.  The gas can only lose angular momentum through collisions, creating a delay before gas can be accreted by the SMBH.  Simulations of accreting gas with vorticity onto a SMBH have shown that Bondi-Hoyle overestimates the accretion rate \citep{Krumholz05}.

Alternative models for SMBH growth have been proposed to try and overcome these problems.  \citet*{Debuhr11} introduced an accretion rate that was dependent on the angular momentum of the gas, building on the previous work of \citet{Hopkins10}.  They set the accretion rate proportional to the mean gas surface density, $\Sigma_{\rm gas}$, the local sound speed squared and inversely proportional to the rotational angular frequency of the gas, $\Omega$, i.e. $\dot{M}_{\rm BH} \propto \Sigma_{\rm gas} c_{\rm s}^2/\Omega$.  While this model accounts for the angular momentum of the gas, it is still accreted onto the SMBH without a delay, such as it would experience in the accretion disc.

In an attempt to account for both the angular momentum and the delayed accretion \citet*[][hereafter PNK11]{Power11} proposed a two stage accretion disc particle model for black hole accretion.  As opposed to approximating the accretion rate based on local gas properties, they defined an accretion radius around the black hole particle and any baryonic material passing inside this is swallowed and added to a subgrid accretion disc.  Material is then allowed to accrete onto the SMBH from the accretion disc over a viscous timescale.

Currently the accretion disc particle model has only been used in idealised disc and major merger simulations \citep[PNK11;][]{Wurster13,Newton13}. In this paper we present the first implementation of this model in a cosmological, large-scale simulation including cooling, star formation and feedback.  In Section \ref{sec:meth} we describe our simulation and give detailed descriptions of the two black hole growth models we have implemented.  In Section \ref{sec:comp} we find the optimal parameters for the accretion disc particle model to reproduce the local black hole density and then compare it to a modified Bondi-Hoyle prescription through mass functions and local scaling relations.  Finally in Section \ref{sec:sum} we summarise our findings from comparing the two growth models and state their suitability to cosmological, large-scale simulations.

\section{Methods}
\label{sec:meth}

The simulations performed in this paper were carried out using a modified version of the $N$-body/SPH code \textsc{gadget}-3 \citep[last described in][]{Springel05}.  The code was modified to include star formation, supernova feedback, radiative cooling, chemodynamics, black hole accretion and AGN feedback.  These were implemented as part of the OverWhelmingly Large Simulation project (OWLS) and are described fully in \citet{Schaye10} and summarised in Section \ref{sec:rad}.  We adopted a flat $\Lambda$-Cold Dark Matter ($\Lambda$CDM) cosmology with parameters: $\{\Omega_m,\Omega_b,\Omega_\Lambda,\sigma_8,n_s,h\}=\{0.238,0.0418,0.762,0.74,0.951,0.73\}$ as determined from the Wilkinson Microwave Anisotropy Probe 3-year results \citep[WMAP-3;][]{Spergel07} and identical to that used by \citet[][hereafter BS09]{Booth09}.

The analysis performed in this paper was conducted on a simulation of a cubical volume of the Universe of side length $50\, h^{-1} \rm Mpc$ comoving, realised using $256^3$ dark matter particles and $256^3$ gas particles giving a dark matter particle mass of $4.06\times10^{8}\, h^{-1} \rm M_{\odot}$ and a gas particle mass of $8.64\times10^{7}\, h^{-1} \rm M_{\odot}$.  The gravitational softening was set to be 0.04 times the mean comoving inter-particle separation down to $z=2.91$, below which a fixed proper scale of $2\, h^{-1} \rm kpc$ was used.  Initial conditions were generated at $z=127$ using the Zel'dovich approximation to linearly evolve positions from an initially glass-like state.  Haloes were found using \textsc{subfind} \citep{Springel01} which produces similar overall results to other halo finders \citep{Muldrew11,Knebe11}.

\subsection{Physics Models}
\label{sec:rad}

In addition to the standard SPH treatment, a number of subgrid models were introduced to represent various physical processes.  A full description of these models can be found in \citet{Schaye10} and references therein, but a summary is given here.

Star formation within the simulation is governed by the method described in \citet{Schaye08}.  Due to the lack of resolution and physical understanding to simulate star formation directly, an effective equation of state is applied for densities $n_{\rm H}>n_{\rm H}^{*}$ where $n_{\rm H}$ is the hydrogen number density and $n_{\rm H}^{*}=0.1\, \rm cm^{-3}$.  The gas is then considered star forming and follows $P \propto \rho^{\gamma}$ with $\gamma=4/3$ and normalised to $P/k=10^{3}\, \rm cm^{-3}\, K$ where $P$ is the pressure and $\rho$ is the density. The gas is then allowed to form stars at a pressure-dependent rate that reproduces the Schmidt-Kennicutt law \citep{Kennicutt98} renormalised to a \citet{Chabrier03} stellar initial mass function (IMF).

Supernovae feedback is modelled kinetically following \citet{DV08}, a variation on the model previously described in \citet{Springel03}.  Energy is injected locally by kicking gas particles into winds and is described by two parameters.  The first is the initial mass-loading, $\eta=\dot{M}_{\rm w} / \dot{M}_{*}$, which is the ratio of the initial amount of gas put into the wind, $\dot{M}_{\rm w}$, compared with the local star formation rate, $\dot{M}_{*}$, and the second is the wind velocity, $v_{\rm wind}$.  Values of $\eta=2$ and $v_{\rm wind}=600\, \rm km\, s^{-1}$ were used in this work which corresponds to 40 percent of the total amount of supernova energy.

Metal enrichment was implemented following \citet{Wiersma09b}.  We follow the timed release of 11 different elements (hydrogen, helium, carbon, nitrogen, oxygen, neon, magnesium, silicon, sulphur, calcium and iron) from massive stars (Type II supernovae and stellar winds) and intermediate-mass stars (Type Ia supernovae and asymptotic giant branch stars), assuming a \citet{Chabrier03} IMF in the range 0.1 to $100\, \rm M_{\odot}$.  Radiative cooling was implemented following \citet{Wiersma09a}.  Net cooling rates were calculated element-by-element in the presence of the cosmic microwave background and a \citet{Haardt01} model for the UV/X-ray background radiation from quasars and galaxies.  The contributions of the 11 elements were interpolated as a function of density, temperature and redshift from tables precomputed by \textsc{cloudy} \citep[last described in][]{Ferland98}, assuming the gas to be optically thin and in (photo-)ionisation equilibrium. 

\subsection{Black Hole Models}
\label{sec:BHMod}

The black hole models considered here can be split into three parts: seeding, growth and feedback.  As part of this investigation we have looked at two different growth models, that of BS09 (Section \ref{sec:BS09}) and that of PNK11 (Section \ref{sec:PNK11}).  The seeding and feedback prescriptions are not varied in this study to allow for a fair comparison of the effects of different growth models.

Black holes were seeded, following the method of \citet{Sijacki07}, using a recursive Friends-of-Friends algorithm \citep[FoF;][]{Davis85} on the dark matter particles.  FoF was run evenly in log expansion factor, $a$, such that $\Delta a=0.02\, a$, which corresponds to $\sim250\, \rm Myr$ ($\sim70\, \rm Myr$) at redshift zero (three).  Dark matter haloes found containing at least 100 particles ($M_{\rm halo,min}=4.06\times10^{10}\, h^{-1} \rm M_{\odot}$) were seeded with a black hole sink particle \citep{Springel05a} if they did not already contain one.  The most gravitationally bound baryonic particle in the halo is converted into a black hole particle with seed mass $10^{-3}M_{\rm gas}$ ($M_{\rm seed}=8.64\times10^{4}\, h^{-1} \rm M_{\odot}$).

Black holes were then left to grow through accretion and mergers following either the BS09 or PNK11 model.  In both cases the accretion rate was Eddington limited:

\begin{equation}
\label{eq:edd}
\dot{M}_{\rm Edd}=\frac{4 \pi G M_{\rm BH} m_{\rm p}}{\epsilon_{\rm r} \sigma_{\rm T} c}
\end{equation}

\noindent where $G$ is the gravitational constant, $M_{\rm BH}$ is the SMBH mass, $m_{\rm p}$ is the proton mass, $\epsilon_{\rm r}$ is the radiative efficiency of a black hole \citep[taken as 0.1 throughout;][]{Shakura73}, $\sigma_{\rm T}$ is the Thomson cross-section for the scattering of free electrons and $c$ is the speed of light.  Black holes are allowed to merge in both accretion models when they pass within a smoothing length (the distance to the 48th nearest gas neighbour), $h_{\rm BH}$, of each other and have a relative velocity smaller than the circular velocity at that distance ($v_{\rm rel}=\sqrt{G M_{\rm BH}/h_{\rm BH}}$).

Feedback from the SMBH is implemented thermally (rise in thermal energy), as opposed to kinetically (rise in kinetic energy) for supernova.  This is the same as the BS09 feedback model, but different to PNK11.  They adopted the model of \citet*{Nayakshin09} where virtual particles are emitted by the SMBH in a Monte-Carlo fashion that interact directly with the SPH density field and deposit their momentum in a region dictated by the optical depth.  The amount of energy released is independent of the environment and no attempt is made to separate the `quasar mode' and `radio mode' feedback.  For each timestep, $\Delta t$, the amount of energy released is:

\begin{equation}
\label{eq:feed}
E_{\rm feed} = \epsilon_{\rm f} \epsilon_{\rm r} \dot{M}_{\rm BH} c^{2} \Delta t
\end{equation}

\noindent where $\epsilon_{\rm f}$ is the efficiency with which a black hole couples the radiated energy into its surroundings.  A value of 0.15 is adopted to produce a good match with observations (BS09).  To ensure that the feedback energy is not immediately radiated away, a minimum heating temperature is imposed and black holes only release energy when they have obtained enough to raise the temperature of $n_{\rm heat}$ particles by $\Delta T_{\rm min}$.  This corresponds to:

\begin{equation}
E_{\rm crit} = \frac{n_{\rm heat} M_{\rm gas} k_{\rm B} \Delta T_{\rm min}}{(\gamma - 1) \mu m_{\rm H}}
\end{equation}

\noindent where $k_{\rm B}$ is the Boltzmann constant, $\gamma=4/3$, $\mu$ is the mean molecular weight (0.58 for a fully ionised gas of primordial composition) and $m_{\rm H}$ is the mass of Hydrogen.  BS09 found that adopting $\Delta T_{\rm min}=10^8\, \rm K$ and $n_{\rm heat}=1$ was sufficient to balance the change in temperature being too small and the timescale between heating being too long.  The energy released by the black hole is then equally distributed into a random fraction $n_{\rm heat}/N_{\rm ngb}$ of the black hole's neighbouring gas particles.

An additional change is made to star forming particles receiving feedback energy.  Particles undergoing star formation are constrained by an effective equation of state, following the \citet{Schaye08} model outlined in Section \ref{sec:rad}, but this is not suitable if they undergo strong heating from the black hole.  Particles that are heated $0.5\, \rm dex$ above the equation of state in a single timestep are removed from the equation of state and are no longer considered star forming.  If their temperature drops at a later time to less than $0.5\, \rm dex$ above the equation of state, they are returned to the equation of state and are considered star forming once more.

\subsubsection{Modified Bondi-Hoyle (BS09)}
\label{sec:BS09}

The BS09 model for black hole growth uses a modified Bondi-Hoyle \citep{Bondi44,Bondi52} prescription to describe the accretion onto black holes. This model builds upon \citet{Springel05a}.  Initially black holes are seeded as described in Section \ref{sec:BHMod}.  The new black hole particle has a SMBH mass corresponding to the seed mass, but the particle mass used in the gravity calculations remains the same as the total mass of the baryonic particle before conversion.  The accretion rate is then calculated as:

\begin{equation}
\dot{M}_{\rm acc}=\alpha \frac{4 \pi G^2 M_{\rm BH}^2 \rho}{(c_{\rm s}^2+v^2)^{3/2}}
\end{equation}

\noindent where $c_{\rm s}$ and $\rho$ are the sound speed and gas density of the local medium, $v$ is the velocity of the black hole relative to the ambient medium and $\alpha$ is a dimensionless efficiency parameter given by:

\begin{equation}
\alpha=\left\{ \begin{array}{cc}
    1 & {\rm if}\,n_{{\rm H}}<n_{{\rm H}}^* \\
    \Big(\frac{n_{{\rm H}}}{n_{{\rm H}}^*}\Big)^\beta &
    \textrm{otherwise}\end{array}\right.
\end{equation}

\noindent where $\beta=2$ (our results are insensitive to $\beta$ for any $\beta>1$; see BS09).  In \citet{Springel05a}, and other works using that method, a value of $\alpha=100$ is adopted.  BS09 argue that for low-density gas such a boost is not justified, as Bondi-Hoyle can be accurately modelled, and choose to boost the accretion rate only in the regime where the Bondi radius is unresolved.  The accretion rate is Eddington limited meaning that it cannot exceed that given by Equation \ref{eq:edd}.  The growth rate of the black hole is then given by:

\begin{equation}
\dot{M}_{\rm BH}=\dot{M}_{\rm acc}(1-\epsilon_{\rm r})
\end{equation}

To account for the accreted mass onto the SMBH, baryonic particles are stochastically swallowed by the black hole particle with probability:

\begin{equation}
 p_i=\left\{ \begin{array}{ll}
    (M_{\rm BH}-M_{{\rm part}})\rho^{-1}W(r_{{\rm BH}}-r_i,h_{{\rm BH}}) & \textrm{if } M_{\rm BH}>M_{{\rm part}} \\
    0                            & \textrm{otherwise}\end{array}\right.
 \end{equation}

\noindent where $W(r_{{\rm BH}}-r_i,h_{{\rm BH}})$ is the SPH kernel evaluated between the black hole and gas particle $i$.  The baryonic particle mass is added to the the black hole particle mass, but no change is made to the SMBH mass.

\subsubsection{Accretion Disc Particle (PNK11)}
\label{sec:PNK11}

The Accretion Disc Particle (ADP) model of PNK11 relies on two free parameters to control the accretion onto the black hole.  An accretion radius, $R_{\rm acc}$, and a viscous timescale, $t_{\rm visc}$. Any baryonic particle that crosses within an accretion radius of the black hole is swallowed and added to an accretion disc.  The SMBH then accretes the disc mass over a viscous timescale giving an accretion rate of:

\begin{equation}
\dot{M}_{\rm BH}={\rm min} \left( \frac{M_{\rm disc}}{t_{\rm visc}},\dot{M}_{\rm Edd} \right)
\end{equation}

\noindent where $M_{\rm disc}$ is the mass in the accretion disc.  The overall black hole particle mass is then given by:

\begin{equation}
M_{\rm part}=M_{\rm BH}+M_{\rm disc}
\end{equation}

\noindent Initially, when seeded, the SMBH is assigned the seed mass and the total baryonic mass minus the SMBH seed is assigned to the accretion disc.  This leaves the total mass of the black hole particle the same as the baryonic particle it was seeded from.

Although designed to be a subgrid model representing a SMBH and its tightly bound accretion disc, it will be shown in Section \ref{sec:comp} that for cosmological simulations the accretion discs produced are too massive to be physical. Therefore, from here on in we will avoid referring to this model as ADP.

\section{Results}
\label{sec:comp}

\begin{figure*}
  \psfrag{mass}[][][1][0]{${\rm log}[M_{\rm BH}/(h^{-1}{\rm M}_{\odot})]$}
  \psfrag{phi}[][][1][0]{${\rm log}[\Phi/(h^{3}{\rm Mpc}^{-3}{\rm dlog(M)})]$}
  \psfrag{a}[l][][1][0]{$R_{\rm acc}$}
  \psfrag{b}[l][][1][0]{$t_{\rm visc}$}
  \includegraphics[width=150mm]{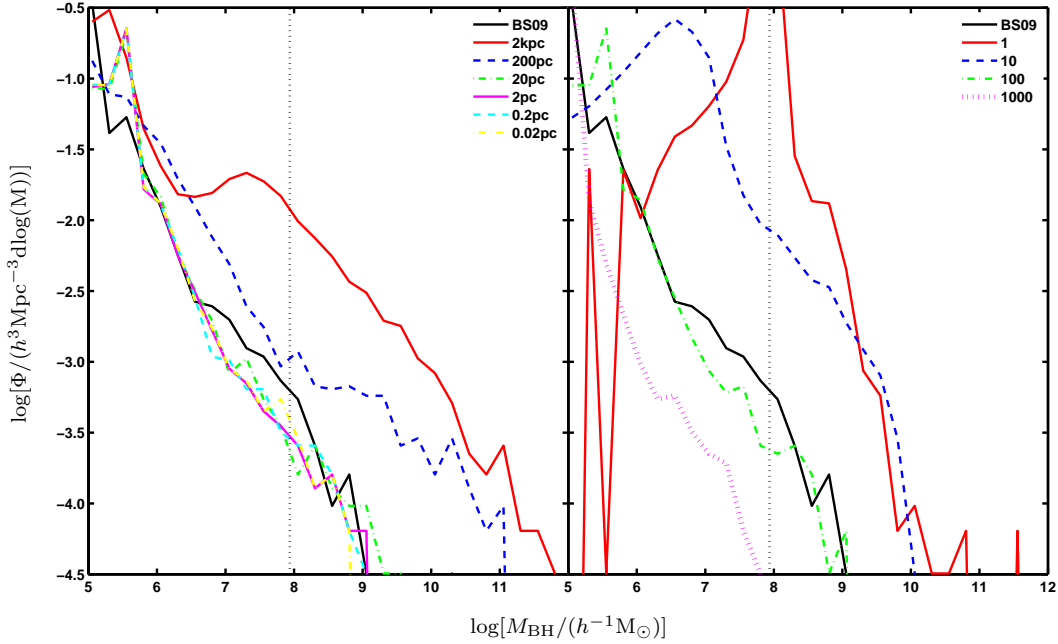}
  \caption{A study of the two free parameters in the PNK11 model. Left: Starting with the accretion radius equal to the gravitational softening and then decreasing by factors of ten for a fixed viscous timescale of $100\, t_{\rm dyn}$. Right: Starting with the viscous timescale equal to the dynamical time and then increasing by factors of ten for a fixed accretion radius of $2\, h^{-1} {\rm pc}$. The black solid line in both panels corresponds to the BS09 model, while the vertical black dotted line is the mass of a single gas particle.}
\label{fig:param}
\end{figure*}

The PNK11 model of black hole growth works using two free parameters, the accretion radius, $R_{\rm acc}$, and the viscous timescale, $t_{\rm visc}$.  To accurately model the growth, these parameters need to be set within the model to reproduce the $z=0$ black hole mass density as closely as possible.  In PNK11 it is suggested that the accretion radius should be set to the smallest resolvable scale of the simulation, of the order the gravitational softening, and the viscous timescale should satisfy $t_{\rm visc} > t_{\rm dyn}(R_{\rm acc})$ where  $t_{\rm dyn}(R_{\rm acc})$ is the dynamical time at the accretion radius.  Using these values as an initial starting point, Figure \ref{fig:param} shows the $z=0$ black hole mass function for various values of the accretion radius with a fixed viscous timescale (left panel) and various values of viscous timescale with a fixed accretion radius (right panel).  The mass function from BS09 is also shown as an illustration.

It is immediately apparent that an accretion radius of the order the gravitation softening ($2\, h^{-1}{\rm kpc}$) is much too large in low resolution cosmological volumes.  This results in a large number of baryonic particles being swallowed, producing overly massive black holes.  This also affects the number of black holes at a given mass, with too many being produced at all values.  This is caused by strong feedback which is triggered by the amount of energy released being related to the accretion rate (Equation \ref{eq:feed}).  For the large number of baryonic particles swallowed, the black hole accretion rate is very high causing large amounts of energy to be released disrupting the structures.  Decreasing the accretion radius by factors of ten shows convergence for values of $20\, h^{-1}{\rm pc}$ ($10^{-2}$ of the gravitational softening) or less.  Within this distance all baryonic particles are accreted by the black hole, regardless of the accretion radius size.

Adopting a value of $2\, h^{-1}{\rm pc}$, which is of the order the physical value used in PNK11 and within the converged values, we then vary the viscous timescale.  Initially a value equal to the dynamical time at the gravitational softening radius is used.  This radius is larger than the accretion radius.  Using such a short viscous timescale causes very rapid accretion that not only produces too massive and too many black holes as seen for large accretion radii, but also produces the wrong shape to the distribution.  Increasing the viscous timescale to 100 times this value produces a black hole mass function similar to BS09 while continuing to increase it produces black holes that are not massive enough.

\begin{figure}
  \psfrag{mass}[][][1][0]{${\rm log}[M_{\rm BH}/(h^{-1}{\rm M}_{\odot})]$}
  \psfrag{phi}[][][1][0]{${\rm log}[\Phi/(h^{3}{\rm Mpc}^{-3}{\rm dlog(M)})]$}
  \includegraphics[width=90mm]{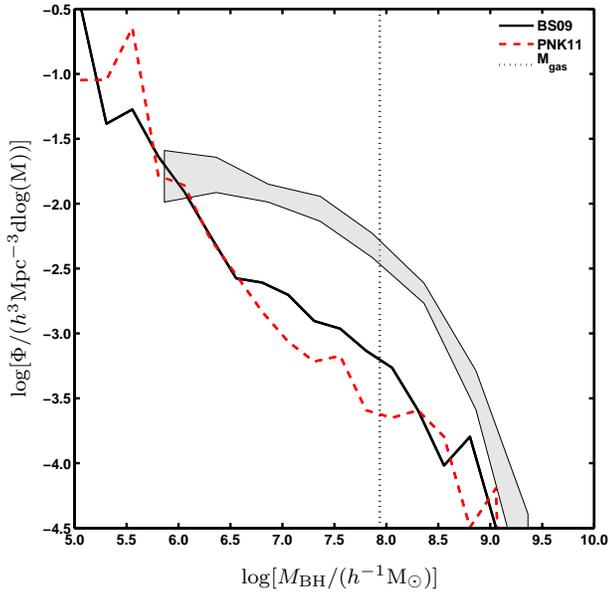}
  \caption{$z=0$ black hole mass function for BS09 (black solid line) and PNK11 (red dashed line) using the best fit parameters from Figure \ref{fig:param}. The shaded region represents the observed mass function taking into account uncertainty from the different methods used to measure it \citep{Shankar09,Kelly12}.  The vertical black dotted line is the mass of a single gas particle.  Both models produce mass function that are too steep and do not tend to follow the Schechter function shape.}
\label{fig:Mfun0}
\end{figure}

\begin{figure}
  \psfrag{mass}[][][1][0]{${\rm log}[M_{*}/(h^{-2}{\rm M}_{\odot})]$}
  \psfrag{phi}[][][1][0]{${\rm log}[\Phi/(h^{3}{\rm Mpc}^{-3}{\rm dlog(M)})]$}
  \includegraphics[width=90mm]{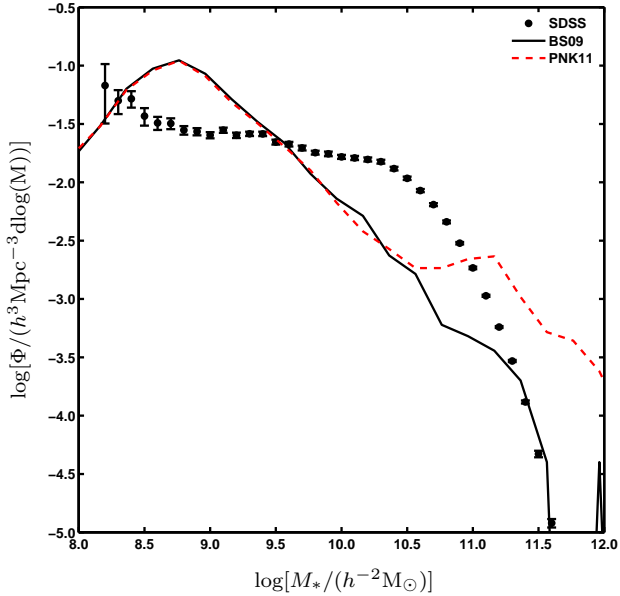}
  \caption{$z=0$ stellar mass function for BS09 (black solid line) and PNK11 (red dashed line). Points correspond to the SDSS mass function from \citet*{Yang09}.  PNK11 has a lower black hole accretion rate to BS09, which weakens the feedback leading to high mass galaxies becoming too massive.}
\label{fig:mgalf}
\end{figure}

From studying the two free parameters of the PNK11 model, values of $R_{\rm acc}=2\, h^{-1}{\rm pc}$ and $t_{\rm visc}=100\, t_{\rm dyn}(R_{\rm soft})$ give the closest black hole mass function to the BS09 model at $z=0$, which was modelled to reproduce the local black hole mass density.  These two mass functions are plotted in Figure \ref{fig:Mfun0} along with the observed uncertainty based on the different methods used to measure it \citep{Shankar09,Kelly12}.  Attempting to measure the mass function observationally is an indirect process.  The mass is inferred through relations with velocity dispersion, stellar mass and spheroid luminosity, and using these different methods leads to the scatter represented by the grey band.  In addition to these, the fundamental plane and S\'{e}rsic index can also be used to measure black hole mass, but these methods underestimate the low mass end relative to the other methods \citep{Shankar09}.  Both models fail to reproduce the observed mass function, underestimating the number of black holes at masses greater than ${\rm log}[M_{\rm BH}/(h^{-1}{\rm M}_{\odot})] \sim 6$.  Below masses of ${\rm log}[M_{\rm BH}/(h^{-1}{\rm M}_{\odot})] \sim 6$ the simulated mass functions continue to rise down to the seed mass, but observational data from \citet{Shankar09} is not available in this range.  Assuming that the mass function stays flat, this would suggest an over production of black holes in this mass range.  Overall the two simulated mass functions do not follow the expected Schechter function shape and are much more linear.  It should be noted that the adopted value of $\sigma_{8}$ is lower in the simulations compared with current observations \citep{Planck13}.  However, this should not effect the shape of the mass function, just the position, and it will be shown in Figure \ref{fig:mgalf} that, despite this, there is no disagreement in the position of the high mass end of the galaxy stellar mass function.

One of the most difficult problems to overcome in modelling black hole growth in simulations of cosmological volumes is that of resolution.  In Figures \ref{fig:param} and \ref{fig:Mfun0} the vertical dotted line represents the mass of a single gas particle ($8.64\times10^{7}\, h^{-1} \rm M_{\odot}$). This is approximately three orders of magnitude larger than the seed mass of the black hole, while at the same time two orders of magnitude smaller than the observed largest SMBHs.  Modelling such a large range of masses is a difficult task when the resolution is roughly in the centre of the mass range, especially for the PNK11 model.  PNK11 was designed to relate the accretion rate more directly to the position of the baryonic particles as opposed to the smoothed density.  This allows for periods of no accretion that is not possible in BS09.  Seeding, let alone accretion, leads to an accretion disc that is massive relative to the SMBH giving a huge fuel supply.  Only once the black holes have grown significantly do they become comparable in mass to the gas particles and these two stages are difficult to combine together.

\begin{figure}
  \psfrag{BH}[][][1][0]{${\rm log}[M_{\rm BH}/(h^{-1}{\rm M}_{\odot})]$}
  \psfrag{disc}[][][1][0]{${\rm log}[M_{\rm disc}/(h^{-1}{\rm M}_{\odot})]$}
  \includegraphics[width=90mm]{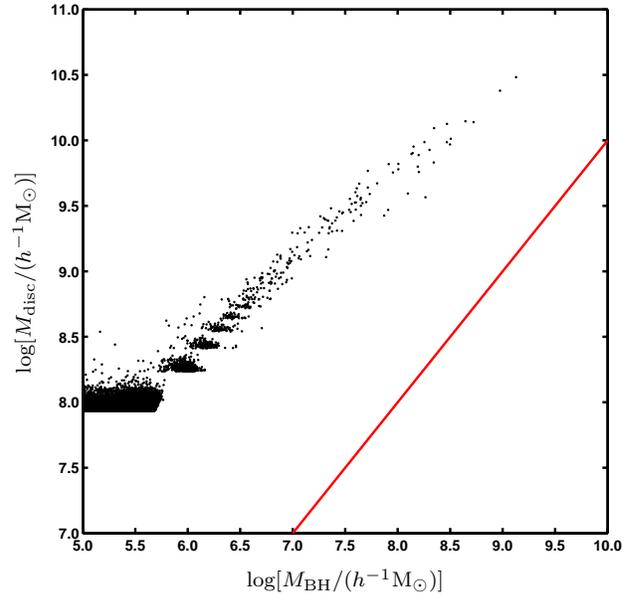}
  \caption{The accretion disc mass ($M_{\rm disc}$) against the black hole mass ($M_{\rm BH}$) for the PNK11 model.  The red line denotes the 1:1 relation.  Black holes in the PNK11 model have accretion discs that are significantly more massive than the black hole, which would lead to instabilities \citep{Thompson05}.}
  \label{fig:disc}
\end{figure}

As mentioned, at early times the accretion disc will be much larger than the SMBH and Figure \ref{fig:disc} shows that this persists to $z=0$ for all masses of black hole.  The solid red line represents the 1:1 line which illustrates how much larger the accretion discs are.  Accretion discs of this size would be unstable, as typically they should be significantly less massive than the black hole \citep{Thompson05}.  Even adopting the two smaller accretion radii from Figure \ref{fig:param} does not change this result.  Accreting gas particles that are three orders of magnitude larger than the seed black hole on a scale that is three orders of magnitude smaller than the gravitational softening leads to too much material being added and the viscous timescale dictating the growth, as opposed to the mass accretion.

\begin{figure*}
  \psfrag{mass}[][][1][0]{${\rm log}[M_{\rm BH}/(h^{-1}{\rm M}_{\odot})]$}
  \psfrag{sigma}[][][1][0]{$\sigma/(\rm km~s^{-1})$}
  \psfrag{a}[l][][1][0]{BS09}
  \psfrag{b}[l][][1][0]{PNK11}
  \includegraphics[width=150mm]{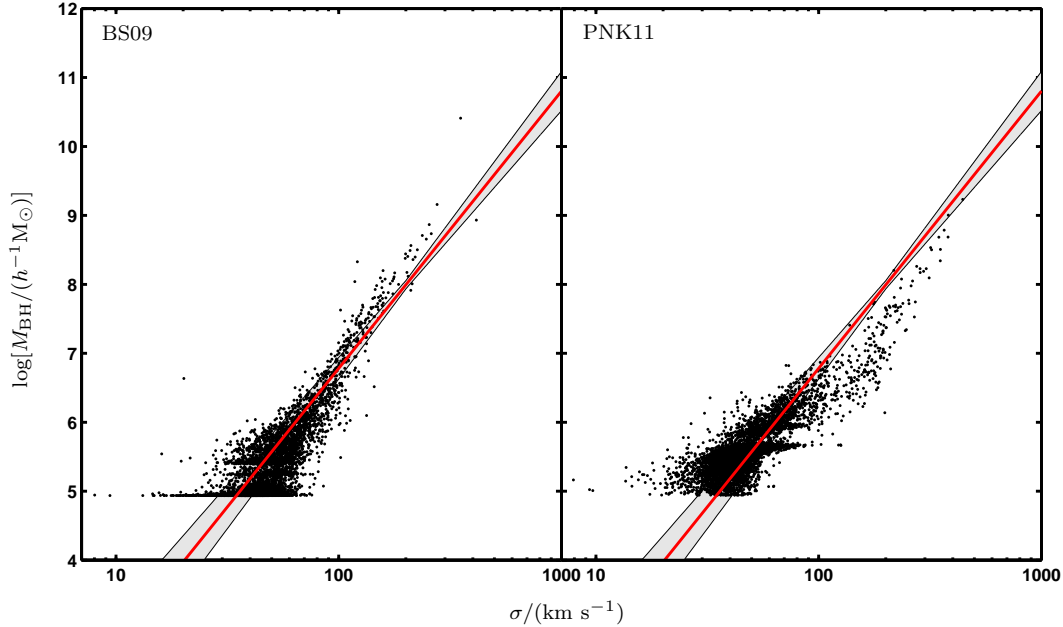}
  \caption{The $M_{\rm BH}-\sigma$ relation for BS09 (left) and PNK11 (right). The red line and shading represents the observed best fit and the uncertainty on this fit from \citet{Tremaine02}.  This is different to the scatter on the data which is larger.  BS09 is well fit by the observations, but PNK11 tends to produce larger velocity dispersions for high mass black holes.}
  \label{fig:sig}
\end{figure*}

A common test of black hole models is their ability to reproduce local scaling relations.  Figure \ref{fig:sig} shows the black hole mass--stellar velocity dispersion ($M_{\rm BH}-\sigma$) relation for BS09 (left panel) and PNK11 (right panel).  The red line in each panel represents the best fit to the observational data from \citet{Tremaine02}, with the shaded region representing the uncertainty on this fit.  The actual scatter in this relation is larger than the shaded region which just represents the range of lines that could be fitted.  Black hole masses correspond to the total black hole mass of the halo as determined by \textsc{subfind}.  BS09 reproduces the $M_{\rm BH}-\sigma$ relation well, although is marginally steeper than the observed relation.  PNK11 fits the relation well for low mass black holes, but for higher masses follows a relation that is offset to higher values of $\sigma$.

\begin{figure*}
  \psfrag{star}[][][1][0]{${\rm log}[M_{*}/(h^{-2}{\rm M}_{\odot})]$}
  \psfrag{bh}[][][1][0]{${\rm log}[M_{\rm BH}/(h^{-1}{\rm M}_{\odot})]$}
  \psfrag{a}[l][][1][0]{BS09}
  \psfrag{b}[l][][1][0]{PNK11}
  \includegraphics[width=150mm]{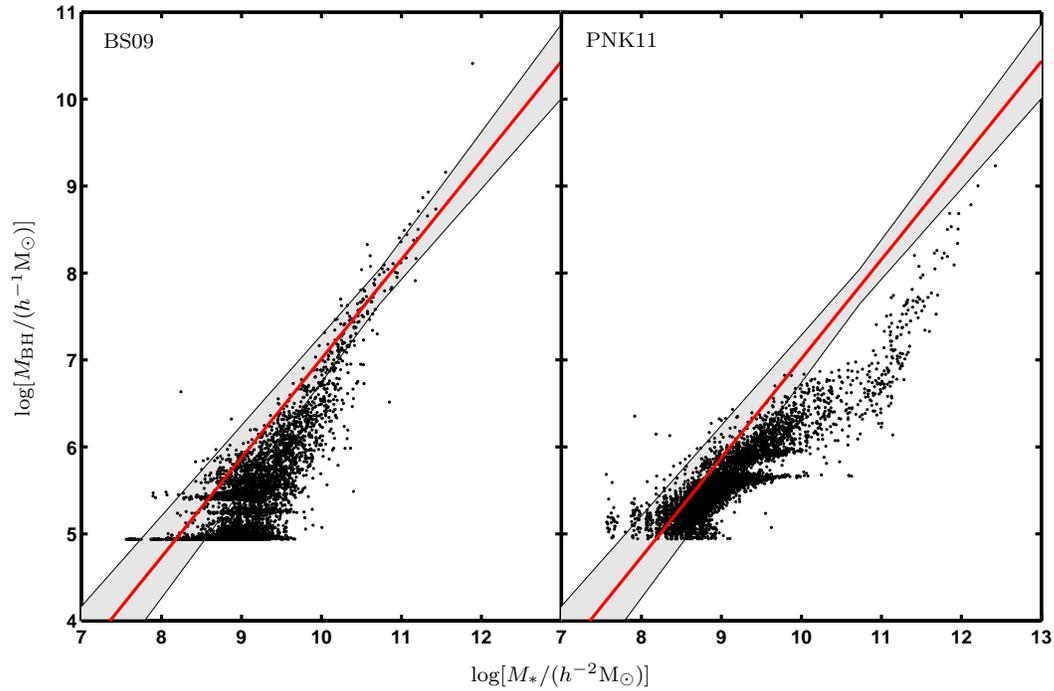}
  \caption{The $M_{\rm BH}-M_{*}$ relation for BS09 (left) and PNK11 (right). The red line and shading represents the observed best fit and the uncertainty on this fit from \citet{Marleau12}.  This is different to the scatter on the data which is larger.  BS09 is well fit by the observations, but tends to produce higher stellar mass galaxies for low mass black holes.  PNK11 tends to produce larger stellar mass galaxies for high mass black holes.}
  \label{fig:star}
\end{figure*}

Another well known scaling relation is that of black hole mass--stellar bulge mass.  \citet{Marleau12} argue that this relation is actually independent of morphology and is really a relation with total stellar mass ($M_{\rm BH}-M_{*}$).  In Figure \ref{fig:star} we plot the $M_{\rm BH}-M_{*}$ relation for BS09 (left panel) and PNK11 (right panel).  This is more accurate than comparisons with bulge mass, as the resolution of our simulations is too low to define bulges or morphology.  The red line in each panel is the best fit to the observational data from \citet{Marleau12} and, again, the shaded region is the uncertainty on this fit, with the scatter of the data being larger.  In general BS09, while close to the relation, is slightly steeper than the observed data.  At higher masses BS09 lies on the observed line, but for low mass black holes the stellar masses tend to be larger than expected.  PNK11 better fits the data for low mass black holes, but still produces galaxies with a slightly higher stellar mass.  For high mass black holes there is again an offset similar to Figure \ref{fig:sig} with the galaxies having a higher stellar mass and following a linear relation.

To better understand the cause of these deviations in the $M_{\rm BH}-\sigma$ and $M_{\rm BH}-M_{*}$ relations, we plot the stellar mass function for the galaxies in Figure \ref{fig:mgalf} and compare it to the observed stellar mass function from the Sloan Digital Sky Survey \citep*[SDSS;][]{Yang09}.  For low mass galaxies the two stellar mass functions are the same, and this is expected as supernova feedback plays a dominant role in shaping the function in this range.  At the high mass end, the PNK11 model produces much more massive galaxies than that of BS09 or that observed.  This is in agreement with Figures \ref{fig:sig} and \ref{fig:star} which show that galaxies with high mass black holes have higher than observed stellar masses and velocity dispersions.  This is down to the feedback proving ineffective from these black holes.  The amount of energy released by the black hole is related to the accretion rate (Equation \ref{eq:feed}) and demonstrates that the accretion rates in PNK11 are lower than those of BS09.  A secondary effect that might weaken the feedback is that the energy released is placed into the neighbours of the black hole particle, which are the closest particles to the black hole and are at risk of accretion. 

\begin{figure}
  \psfrag{z}[][][1][0]{$z$}
  \psfrag{Density}[][][1][0]{$\rho_{\rm BH}/(h^2{\rm M}_{\odot}{\rm Mpc}^{-3})$}
  \includegraphics[width=90mm]{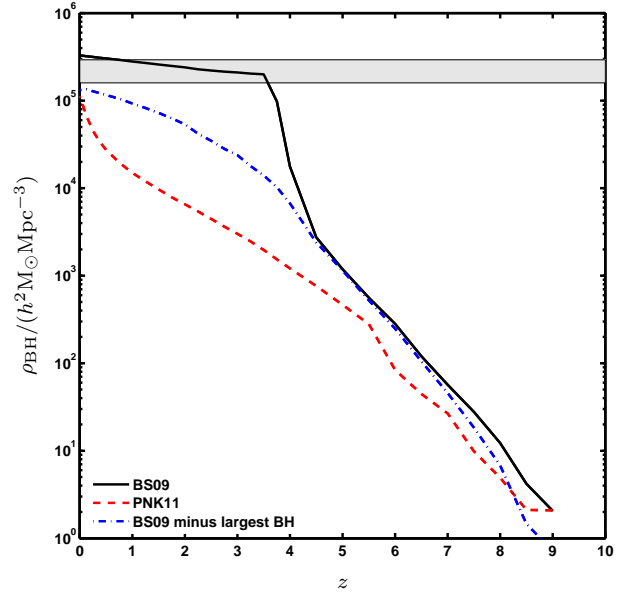}
  \caption{The evolution of the black hole mass density with redshift for BS09 (black solid line), PNK11 (red dashed line) and BS09 less the mass of the largest black hole in the volume (blue dot-dashed line).  The shaded region corresponds to the observed black hole mass density at $z=0$ from \citet{Shankar09}. The PNK11 model predicts a smooth growth in the black hole mass density, while BS09 undergoes three different regimes dominated by the growth of the largest black hole in the volume.  Subtracting this yields a smoother growth distribution steeper than PNK11.}
  \label{fig:evo}
\end{figure}

Finally we consider the evolution of the black hole mass density in Figure \ref{fig:evo}.  The grey band represents the observed $z=0$ density from \citet{Shankar09} and BS09 slightly overestimates this value.  The model of PNK11 produces a density that is a factor of three smaller than BS09 at $z=0$ and also outside the observed value.  Overall, PNK11 has a smooth evolution of the density with redshift.  BS09 has a less smooth distribution and grows in three stages.  Firstly there is a smooth growth that is steeper than PNK11, before a sudden rapid phase that then flattens out.  This period of small change in the density at low redshift is consistent with downsizing.  Although the mass functions looked similar in Figure \ref{fig:Mfun0}, the density appears very different.  This is down to the growth of one very massive black hole ($2.47\times10^{10}\, h^{-1} \rm M_{\odot}$ at $z=0$) that is not present for the PNK11 model, and is disconnected by over an order of magnitude from the second largest and so is not shown in Figure \ref{fig:Mfun0}.  Subtracting this black hole from the volume and recalculating the density yields a smoother evolution that agrees with PNK11 at $z=0$.  The steeper growth in BS09 before flattening means that the accretion rates will be higher at high redshift making feedback more effective in this regime compared with PNK11, preventing the over production of massive galaxies (Figure \ref{fig:mgalf}).

\section{Summary and Conclusions}
\label{sec:sum}

SMBHs are known to play an important role in galaxy evolution and a number of properties are strongly correlated with their mass.  To produce the most realistic models of galaxies, the black hole growth also needs to be modelled accurately.  We have implemented the PNK11 accretion disc model of black hole growth into a large-scale, cosmological simulation including star formation and feedback, and compared it with a modified Bondi-Hoyle model of BS09.  Whereas BS09 relates the accretion rate to the local density and sound speed of the gas, $\dot{M}_{\rm BH} \propto \dot{M}_{\rm Bondi} \propto M_{\rm BH}^2\rho/c_{\rm s}^3$, PNK11 uses two free parameters to govern accretion.  Baryonic particles that pass within a given radius are swallowed by the black hole and added to a subgrid accretion disc.  The black hole then accretes this material over a given timescale.

Setting these parameters is an important and non-trivial task to make sure that the $z=0$ black hole masses reflect those observed.  Taking the accretion radius to be equal to the gravitational softening, the smallest resolvable scale, produces black holes that are too massive by the present day.  This is down to the low resolution large-scale, cosmological simulations are currently run at, due to the limitations of computing power.  Below 0.01 times the gravitational softening, the mass functions converge on the same value.  This result is unexpected, as decreasing the radius further would na\"{i}vely suggest less material would be accreted.  Beneath the gravitational softening there is a radius at which all particles will be accreted by the black hole particle.  The conclusions from this is that the accretion radius must be set to a physical size as opposed to relating it to properties of the simulation.  A value of a few parsec is consistent with PNK11, \citet{Wurster13} and this work.

The viscous timescale is a harder parameter to set, as every change in its value produces a different result.  Here we have modified it from previous studies by introducing a black hole mass dependence through the dynamical time.  Other works have stuck to a fixed value.  While the viscous timescale is designed to delay accretion onto the black hole, in this work it has principally been used to buffer the excessive accretion.  One of the advantages of the PNK11 model is that it includes a subgrid accretion disc, but these are too massive to be realistic even for the smallest accretion radii.  The excessive accretion observed has been prevented from reaching the black hole by using very long viscous timescales.

Within this paper we have implemented the model of PNK11 in its simplest form, but further extensions are required for the current resolution of cosmological simulations.  Accretion discs of the scale produced here would fragment leading to star formation, which in turn would lead to further feedback.  This would affect the amount of material available to the black hole, lowering the accretion rate.  Efforts have been made by \citet{Newton13} to improve the subgrid modelling of PNK11 by adding a two stage process.  They accrete gas on the scale of the gravitational softening, but then delay its addition to the accretion disc representing the subgrid behaviour between the gravitational softening and the black hole accretion radius.  The primary advantage of using the PNK11 model is that accretion is measured directly as opposed to being approximated as in the Bondi-Hoyle model.  One possible route to improving the PNK11 model in cosmological simulations is to study the model using high resolution zooms where it has been shown to be effective.  By measuring the accretion at high resolution, using a physically motivated model, an improved subgrid model for low resolution runs can be developed.

Common tests to check the effectiveness of black hole models include comparing with the local mass density and local scaling relations, such as black hole mass--stellar velocity dispersion, $M_{\rm BH}-\sigma$, and black hole mass--total stellar mass, $M_{\rm BH}-M_{*}$.  Both models came close to reproducing these, with only small deviations at the high mass end related to ineffective feedback in the case of PNK11.  For the local black hole density, PNK11 produces a value that is three times smaller than BS09, although for BS09 the total mass is dominated by one very massive black hole ($>10^{10}\, h^{-1}{\rm M}_{\odot}$).  Removing this produces a density similar to PNK11.  Testing black hole models against these relations can be misleading, as deviations from these can be small compared with observational scatter and reproducing these relations does not guarantee that the right mass distribution of black holes is being produced.

An additional test of black hole mass models is to compare with the black hole mass function.  This has been measured in a number of different ways observationally and the uncertainty on it is now well constrained.  Although both models produce mass functions that are similar to each other, and reproduce scaling relations, neither agree with the observed values.  The modelled mass functions do not follow the Schechter function shape, producing steep lines that overestimate the number of black holes at the low mass end, while underestimating the intermediate and high mass end.  \citet{Booth10} demonstrate that $M_{\rm BH} \propto \epsilon_{\rm f}^{-1}$, which means decreasing $\epsilon_{\rm f}$ will increase the masses of the black holes. However, making this change also affects the $M_{\rm BH}-\sigma$ relation, altering the normalisation, and so no longer agrees with observations.  Changing the value of $\epsilon_{\rm f}$ also does not improve the shape of the black hole mass function. The continued rise at low masses may be the result of seeding model, leading to black holes tracing the dark matter halo mass function closer than the galaxy stellar mass function.  Meanwhile the deficit shown at intermediate masses may correspond to the same deficit shown in the galaxy stellar mass function (Figure \ref{fig:mgalf}).  Further work is needed on black hole modelling to address this discrepancy to make sure that the distribution is correct, which in turn will help improve the galaxy stellar mass function.

The most important function of black holes in galaxy formation simulations is the feedback they provide to prevent the formation of overly massive galaxies.  Using the same thermal feedback model of BS09 on PNK11 proves ineffective and the black holes cannot prevent this happening.  Combining this with the large amount of material that is swallowed into the accretion disc suggests that a stronger feedback mechanism is need.  A kinetic regime has the advantage of being able to drive gas particles away from the black hole, preventing this over accretion and may reduce the risk of a particle that receives feedback energy being accreted.  In future work we will look to implement this, as it is apparent that the growth of the black hole and the feedback are strongly coupled and should be treated as one process.  Ideally, a better physical understanding of how the feedback energy from the black hole couples with the surrounding gas needs to be determined in order to improve the implementation within models.

The ability to reproduce local scaling relations has been used to show the success of black hole modelling, but recently \citet{vdB12} have  presented a number of galaxies that do not obey these, containing very massive black holes.  One such example is NGC1277, which has a stellar mass of $1.2\times10^{11}\, {\rm M}_{\odot}$ and a black hole mass of $1.7\times10^{10}\, {\rm M}_{\odot}$.  These galaxies are not constrained by environment and can be found in and out of clusters.  Possible formation channels include some run away process that allows the black hole to accrete gas heavily at high redshift or the possible accretion of star clusters might accelerate growth.  For the BS09 model, we have one case of a very massive black hole and another that is large for its stellar mass.  Until the space density of these objects is better understood, it is unclear at the present time whether these objects fit in with our current models of black hole growth or whether further consideration is needed.

Overall we have demonstrated that modelling the growth and feedback from supermassive black holes is still very much a subgrid process in large-scale, cosmological simulations.  Directly linking the accretion of gas particles to the black hole produces excessive accretion not seen in higher resolution implementations.  Whereas the modified Bondi-Hoyle model reproduces the black hole properties better, this is just an approximation of the accretion rate.  Ideally through modelling we want to say something about the physical processes causing accretion and feedback, but at the current resolution of large-scale, cosmological simulations this is not possible.

\section*{Acknowledgments}

The authors wish to thank Craig Booth and Joop Schaye for useful comments and for supplying their accretion scheme.  We also thank Brandon Kelly for supplying the observational data in Figure \ref{fig:Mfun0} and Sergei Nayakshin and Hanni Lux for useful discussions.  SIM acknowledges the support of the STFC Studentship Enhancement Programme (STEP).  The simulations in this paper were performed on the High Performance Computing (HPC) facilities at the University of Nottingham (www.nottingham.ac.uk/hpc).  Part of the research presented in this paper was undertaken as part of the Survey Simulation Pipeline (SSimPL; http://ssimpl-universe.tk).

\bibliographystyle{mn2e}
\bibliography{bh}

\bsp

\label{lastpage}

\end{document}